\documentclass[10pt,english,abstracton]{scrartcl}
\usepackage[T1]{fontenc}
\usepackage[latin1]{inputenc}
\usepackage{geometry}
\usepackage{amsmath}
\usepackage{graphicx}
\usepackage{amssymb}

\makeatletter
\usepackage{cite}
\usepackage{lastpage}
\usepackage{ae,aecompl}

\usepackage{babel}
\makeatother
\begin{document}

\titlehead{\hfill{}TUM-HEP-575/05}

\title{{\Large The Running of the Cosmological and the Newton}\\
{\Large Constant controlled by the Cosmological Event Horizon}}

\author{{\large Florian Bauer}%
\footnote{Email: \texttt{fbauer@ph.tum.de}%
}{\large }\\
{\large }\\
\emph{\large Physik-Department, Technische Universität München,}\\
\emph{\large James-Franck-Straße, D-85748 Garching, Germany}}

\date{{\large 26 January 2005}}

\maketitle
\begin{abstract}
We study the renormalisation group running of the cosmological and
the Newton constant, where the renormalisation scale is given by the
inverse of the radius of the cosmological event horizon. In this framework,
we discuss the future evolution of the universe, where we find stable
de~Sitter solutions, but also {}``big crunch''-like and {}``big
rip''-like events, depending on the choice of the parameters in the
model.
\end{abstract}
\renewcommand{\textrm}{\text}






\newcommand{\kl}[1]{{\textstyle #1}}

\section{Introduction}

The recently observed \cite{SN98RiessPerl} accelerated expansion
of the universe may have its reason in the existence of dark energy~(DE),
an energy form with negative pressure, which is so far not understood.
The cosmological constant~(CC) is the theoretically simplest candidate
for DE, because it occurs as a classical parameter in Einstein's equations,
and further it has an origin as vacuum energy in quantum field theory~(QFT).
On the other hand, it is difficult to explain the tiny value of the
CC and the actual coincidence of the energy densities of the CC and
non-relativistic matter \cite{CCProblem}.

There are lots of models, which describe DE as a dynamical quantity,
e.g. by using scalar fields. Another possibility is the modification
of the theory of gravity by introducing extra terms in the equations
of the cosmological evolution, or extending our space-time by additional
space-time dimensions. However, in most of these models the accelerated
cosmic expansion is due to new and unknown physics, which often means
a high amount of arbitrariness and limited predictability. 

In this work, we investigate the CC in the sense that it emerges anyway
on a formal level in QFT. There, the zero-point or vacuum energy of
a quantised field has the same equation of state as the CC occurring
in Einstein's equations. Unfortunately, it is unknown how to calculate
its value in a unique way, because it can be written in the form of
a quartically divergent momentum integral like~$\int\textrm{d}^{3}p\cdot p$.
The naive assumption of an ultra-violet (UV) cutoff in this integral
at some known energy scale usually leads to an unobserved high value
of the CC, which is also called the old CC problem. However, the procedure
of renormalisation of (coupling) constants in QFTs can handle infinities,
thereby leading to a dependence of the renormalised constants on some
energy scale~$\mu$. In many cases, this renormalisation scale can
be identified with an external momentum, or at least with some characteristic
scale (e.g. the temperature) of the environment. Studying QFT on curved
space-time~\cite{QFT-curved} leads to infinities in the vacuum expectation
values (VEV) of the energy-momentum tensors of the fields. This can
be treated by renormalisation to yield a scale-dependent or running
CC and a running Newton constant~(NC). The absolute values are still
not calculable, but the change with respect to the renormalisation
scale can be described by the renormalisation group equations~(RGE).
Unlike the running coupling constants in the standard model of particles,
here, the physical meaning of the scale is not given by the theory.
This becomes obvious since, in the language of Feynman graphs, the
formally infinite value of the CC corresponds to a closed loop without
external legs and hence no distinct energy scale. Connecting this
renormalisation scale with physics thus requires an additional theoretical
input. Usual choices in the literature are the Hubble scale, the square
root of the Ricci scalar, and combinations of similar quantities.
The RG running of the CC and the NC has been studied in several different
frameworks and models, some recent results can be found in Refs.~\cite{RGE-1,BetaFkt,RGE-2,RGE-3,RGE-4}.

In our investigation of the RG running of the CC and the NC, we choose
the inverse of the radius of the cosmological event horizon as the
renormalisation scale. Such a horizon usually exists in accelerating
universes like ours. In addition, the possible relevance of the horizon
scale for DE has often been pointed out~\cite{EventHor-2,EventHor-3,EventHor-4}.
In section~\ref{sec:RGEs}, we derive the RGEs and their dependence
on the renormalisation scale and the parameters, mainly the field
masses. Since the event horizon in an evolving universe is not constant
in time, the CC and the NC are also time-dependent, implying that
the usual covariant conservation equations for the energy momentum
tensor have to be modified (section~\ref{sec:EvolScaleFactor}).
In section~\ref{sec:DiscussAnalyt}, we discuss the properties of
the new evolution equation for the cosmic scale factor and derive
some conditions on the existence and the stability of the final states
of the universe. The cosmic fate is the main point of the discussion,
since the characteristic behaviour in the far future depends crucially
on the parameters in the RGEs. In section~\ref{sec:DiscussNumeric},
we illustrate the possible final states of the universe by showing
some numerical solutions and their dependence on the parameters. Finally,
section~\ref{sec:Conclusions} contains our conclusions and some
open points of this setup.

\section{\label{sec:RGEs}Renormalisation group equations}

To formulate the RGEs for the CC and the NC~$G$, we consider free
quantum fields on a curved space-time, namely a Friedmann-Robertson-Walker~(FRW)
universe with a positive CC~$\lambda$. For one fermionic and one
bosonic degree of freedom with masses~$m_{\textrm{F}}$ and~$m_{\textrm{B}}$,
respectively, the $1$-loop effective action can be written in the
form~\cite{QFT-curved}\begin{equation}
S_{\textrm{eff}}=\int\textrm{d}^{4}x\sqrt{-g}\left[\frac{\textrm{Ric}}{16\pi G}-\Lambda+\left(D+\ln\frac{m_{\textrm{F/B}}}{\mu}\right)\cdot\left(\frac{m_{\textrm{F}}^{4}-m_{\textrm{B}}^{4}}{32\pi^{2}}-\frac{\textrm{Ric}}{16\pi^{2}}\left[(\xi-\kl{\frac{1}{6}})m_{\textrm{B}}^{2}-\kl{\frac{1}{12}}m_{\textrm{F}}^{2}\right]\right)\right]+C,\label{eq:Seff}\end{equation}
where~$D=\kl{\frac{1}{2}}\gamma_{\textrm{Euler}}+\lim_{n\rightarrow4}(n-4)^{-1}$
is a divergent term, which does not depend on the renormalisation
scale~$\mu$. Furthermore,~$\xi$ is a coupling constant%
\footnote{In the action~$S=\int\textrm{d}^{4}x\sqrt{-g}\left[\frac{\textrm{Ric}-2\lambda}{16\pi G}+\frac{1}{2}\phi_{;\alpha}\phi^{;\alpha}-\frac{1}{2}[m^{2}+\xi\textrm{Ric}]\phi^{2}\right]$
of a scalar field~$\phi$ on a curved space-time, the constant~$\xi$
occurs in the coupling term~$\xi\cdot\textrm{Ric}\cdot\phi^{2}$
between the scalar field and the Ricci scalar~$\textrm{Ric}$.%
}, and the variable~$C$ represents all further terms in the effective
action, that are neither proportional to the Ricci scalar~$\textrm{Ric}$
nor to the vacuum energy density~$\Lambda:=\lambda/(8\pi G)$.

The relevant $\beta$-functions in the $\overline{\textrm{MS}}$-scheme
for the vacuum energy density~$\Lambda$ and the NC~$G$ are obtained
by the requirement that the effective action~$S_{\textrm{eff}}$
must not depend on the renormalisation scale~$\mu$,\[
\mu\frac{\textrm{d}S_{\textrm{eff}}}{\textrm{d}\mu}=0.\]
Because of this condition, $\Lambda$ and~$G$ have to be treated
as $\mu$-dependent functions in Eq.~(\ref{eq:Seff}), which have
to obey the RGEs given by\[
\mu\frac{\textrm{d}\Lambda}{\textrm{d}\mu}=-\frac{m_{\textrm{F}}^{4}-m_{\textrm{B}}^{4}}{32\pi^{2}},\,\,\,\,\,\mu\frac{\textrm{d}}{\textrm{d}\mu}\left(\frac{1}{G}\right)=-\frac{1}{\pi}\left[(\xi-\kl{\frac{1}{6}})m_{\textrm{B}}^{2}-\kl{\frac{1}{12}}m_{\textrm{F}}^{2}\right].\]
Note, that the divergent term~$D$ has dropped out, leaving over
just the masses~$m_{\textrm{F/B}}$ and~$\xi$. Assuming constant
masses, the RGEs can be integrated. Hence, the equation for the vacuum
energy density reads\begin{equation}
\Lambda(\mu)=\Lambda_{0}(1-q_{1}\ln\kl{\frac{\mu}{\mu_{0}}})\,,\,\,\,\,\Lambda_{0}:=\Lambda(\mu_{0}),\label{eq:RGE-Lambda}\end{equation}
where the sign of the parameter\begin{equation}
q_{1}:=\frac{1}{32\pi^{2}\Lambda_{0}}\left(m_{\textrm{F}}^{4}-m_{\textrm{B}}^{4}\right)\label{eq:q1Def}\end{equation}
depends on whether bosons or fermions dominate. In this context, a
real scalar field counts as one bosonic degree of freedom, and a Dirac
field as four fermionic ones. The generalisation to more than one
quantum field in the RGE, can be achieved by summing over the fourth
powers of their masses. For the NC~$G$ we obtain the RGE in the
integrated form\begin{equation}
G(\mu)=\frac{G_{0}}{1-q_{2}\ln\frac{\mu}{\mu_{0}}}\,,\,\,\,\, G_{0}:=G(\mu_{0}).\label{eq:RGE-G}\end{equation}
Again, we omit the generalisation to more fields, that follows from
summing over the squared masses of the fields. For one bosonic and
one fermionic degree of freedom the mass parameter~$q_{2}$ is given
by\begin{equation}
q_{2}:=\frac{G_{0}}{\pi}\left[(\xi-\kl{\frac{1}{6}})m_{\textrm{B}}^{2}-\kl{\frac{1}{12}}m_{\textrm{F}}^{2}\right].\label{eq:q2Def}\end{equation}
Finally, we remark, that Eq.~(\ref{eq:RGE-Lambda}) for the running
vacuum energy density~$\Lambda(\mu)$ was derived in a renormalisation
scheme, which is usually associated with the high energy regime. Unfortunately,
the corresponding covariantly derived equations for the low energy
sector are not known yet~\cite{DecLamG}. Therefore, we prefer to
work with the above RGEs, which were derived in a covariant way, and
study the consequences and the constraints on the mass parameters~$q_{1}$
and~$q_{2}$.

Next, we choose the renormalisation scale~$\mu$ to be the inverse
of the radius~$R$ of the cosmological event horizon. In the FRW
universe the (radial) horizon radius~$R$ at the cosmic time~$t$
is given by\begin{equation}
\mu^{-1}=R(t):=a(t)\cdot\int_{t}^{\infty}\frac{\textrm{d}t^{\prime}}{a(t^{\prime})},\label{eq:EventHor}\end{equation}
where~$a(t)$ is the cosmic scale factor, corresponding to the line
element%
\footnote{The constant~$k$ fixes the spatial curvature of the universe.%
}\[
\textrm{d}s^{2}=\textrm{d}t^{2}-a^{2}(t)\left(\frac{\textrm{d}r^{2}}{1-kr^{2}}+r^{2}\textrm{d}\Omega^{2}\right).\]
For universes, which end within finite time, the upper limit of the
integral in Eq.~(\ref{eq:EventHor}) should be replaced by the time
when the universe ends.

The choice of the scale~$\mu=R^{-1}$ can be motivated by the thermodynamical
properties of the cosmological event horizon. This horizon emits radiation,
whose temperature is given by the Gibbons-Hawking~\cite{EventHor-1}
temperature~$T_{\textrm{GH}}=(2\pi R)^{-1}$, that is proportional
to our renormalisation scale~$\mu$. For a comoving observer in a
de~Sitter universe the only cosmological energy scale is given by
this temperature. We have to admit that this no prove of the rightness
of this choice. On the other hand, the investigation of the cosmological
evolution with this specific renormalisation scale is the main point
of this work, and the resulting solutions are quite interesting. From
Eq.~(\ref{eq:EventHor}) we see that in an evolving universe the
event horizon radius~$R$ and thus the scale~$\mu$ are usually
not constant in time. Therefore, the vacuum energy density~$\Lambda$
and the NC~$G$ will be time-dependent, too. This requires a generalisation
of the covariant conservation conditions, and complicates the solutions
of Friedmann's equations.

\section{\label{sec:EvolScaleFactor}Evolution equation for the scale factor}

In this section, we derive the evolution equation for the cosmic scale
factor~$a(t)$ in the framework of the spatially isotropic and homogeneous
FRW universe with a time-dependent CC and NC. On this background,
radiation and pressureless matter (dust) can both be described by
a perfect fluid with the energy density~$\rho$ and the pressure~$p=\omega\rho$,
where the constant~$\omega$ characterises the equation of state%
\footnote{Dust:~$\omega=0$; radiation:~$\omega=\frac{1}{3}$.%
}. The corresponding energy-momentum tensor for these energy forms
reads\[
T^{\alpha\beta}=(\rho+p)u^{\alpha}u^{\beta}-pg^{\alpha\beta},\]
with~$u^{\alpha}$ being the four-velocity vector field of the fluid.
With our choice of the renormalisation scale,~$G$ and~$\Lambda$
depend only on the cosmic time~$t$. From Einstein's equations\[
G^{\alpha\beta}=8\pi G(\Lambda g^{\alpha\beta}+T^{\alpha\beta})\]
and from the contracted Bianchi identities~$G_{\,\,\,\,\,\,;\beta}^{\alpha\beta}=0$
for the Einstein tensor~$G^{\alpha\beta}$, we obtain the generalised
conservation equations%
\footnote{We do not assume~$T_{\,\,\,\,\,\,;\beta}^{\alpha\beta}=0$.%
}\[
0=[G\Lambda g^{\alpha\beta}+GT^{\alpha\beta}]_{;\beta}\,\,\stackrel{\alpha=0}{=}\,\,\dot{G}(\Lambda+\rho)+G(\dot{\Lambda}+\dot{\rho}+3\kl{\frac{\dot{a}}{a}}\rho(1+\omega)).\]
Note that the simple scaling rule~$\rho\propto a^{-3(1+\omega)}$
for the matter content is not valid anymore, because it is now possible
to transfer energy between the matter and the vacuum, in addition
to~$\dot{G}\neq0$. Therefore, we have to combine the Friedmann equations
for the Hubble scale~$H:=\frac{\dot{a}}{a}$ and the acceleration~$\frac{\ddot{a}}{a}$,\begin{eqnarray}
\left(\frac{\dot{a}}{a}\right)^{2}+\frac{k}{a^{2}} & = & \frac{8\pi}{3}G(t)(\Lambda(t)+\rho(t)),\label{eq:Friedmann-1}\\
\frac{\ddot{a}}{a} & = & \frac{8\pi}{3}G(t)(\Lambda(t)-Q\rho(t)),\,\,\,\, Q:=\frac{1}{2}(1+3\omega),\label{eq:Friedmann-2}\end{eqnarray}
to eliminate the matter energy density~$\rho$. The left-hand side
of the result is abbreviated by~$F(t)$:\begin{equation}
F(t):=\frac{\ddot{a}}{a}+Q\left[\left(\frac{\dot{a}}{a}\right)^{2}+\frac{k}{a^{2}}\right]=\frac{8\pi}{3}G\Lambda\cdot(1+Q).\label{eq:Friedmann-3}\end{equation}
Now the RGEs for~$\Lambda$ and~$G$ from Eqs.~(\ref{eq:RGE-Lambda})
and~(\ref{eq:RGE-G}) are inserted in Eq.~(\ref{eq:Friedmann-3})
to yield, with the specific choice of our renormalisation scale~$\mu=1/R$,
the main equation of this work:\begin{equation}
\frac{\mu_{0}}{\mu(t)}=\frac{R(t)}{R_{0}}=\exp\left[\frac{K_{0}F(t)-1}{q_{1}-q_{2}K_{0}F(t)}\right].\label{eq:Hptglg}\end{equation}
In this equation the constant~$K_{0}$ is defined as\begin{equation}
K_{0}:=\frac{3}{8\pi G_{0}\Lambda_{0}(Q+1)}=\frac{H_{0}^{-2}}{\Omega_{\Lambda0}(Q+1)},\label{eq:K0Def}\end{equation}
where~$\Omega_{\Lambda0}=8\pi G_{0}\Lambda_{0}/(3H_{0}^{2})$ is
the relative vacuum energy density and~$H_{0}$ the Hubble scale
at the time~$t=t_{0}$. 

Note that for a dominant matter energy density~$\rho\gg\Lambda$
and flat spatial curvature~($k=0$), the acceleration quantity~$q:=\frac{\ddot{a}a}{\dot{a}^{2}}$
is given by the negative value of the (new) equation of state parameter~$Q=(1+3\omega)/2$,
which we have introduced in Eq.~(\ref{eq:Friedmann-2}).

\section{\label{sec:DiscussAnalyt}Discussion}

This section is devoted to the discussion of the properties of our
main equation~(\ref{eq:Hptglg}), thereby placing special interest
in the late-time behaviour of the scale factor~$a(t)$. Solving~Eq.~(\ref{eq:Hptglg})
for~$F$ leads to\begin{equation}
K_{0}F=\frac{1+q_{1}\ln\frac{R}{R_{0}}}{1+q_{2}\ln\frac{R}{R_{0}}},\label{eq:K0FofR}\end{equation}
 which is plotted as a function of the horizon radius~$R$ in Figs.~\ref{cap:q1minus2-q2minus01},~\ref{cap:q1minus2-q2plus01},~\ref{cap:q1plus2-q2minus01}
and~\ref{cap:q1plus2-q2plus01} for four cases, depending on the
signs of~$q_{1}$ and~$q_{2}$. For reasonable field masses~$m$,
the magnitude of the mass parameter~$q_{1}$ is always greater than
that of~$q_{2}$. In the definition of the parameter~$q_{2}$ the
field masses~$m$ are divided by today's Planck mass~$M_{\textrm{Planck},0}=1/\sqrt{G_{0}}$,
therefore,~$\left|q_{2}\right|$ is a tiny quantity today. This implies
a very weak running of the NC~$G$, which agrees with the strong
bounds on the time-variation of~$G$~\cite{Gdot}. Additionally,
this has the advantage, that today we are far away from the Landau
pole of~$G(\mu)$, where the function~$F$ diverges. Very high values
of~$\left|F\right|$ could render our calculations invalid, since
we do not implement higher powers of the curvature scalar in the gravitational
action. These contributions are probably relevant in regimes with
large~$\left|F\right|$. Whether they are able to prevent the singular
behaviour in the scale factor, that occurs in the numerical solutions
for some cases, requires further investigation and is not treated
in this paper.

Much less problematic are final states of the universe, that are de~Sitter-like.
In this case, the scale factor acquires the form~$a(t)\propto\exp(H_{\textrm{e}}t)$
for large~$t$, where~$H_{\textrm{e}}$ denotes the constant Hubble
scale, and the radius of the event horizon is given by~$R_{\textrm{e}}=1/H_{\textrm{e}}$.
Plugging this asymptotic form for~$a(t)$ in Eq.~(\ref{eq:K0FofR}),
one arrives at\[
\frac{K_{0}(1+Q)}{R_{\textrm{e}}^{2}}=q_{3}x^{-2}=\frac{1+q_{1}\ln x}{1+q_{2}\ln x},\]
where the variables~$x:=R_{\textrm{e}}/R_{0}$ and~$q_{3}:=K_{0}(1+Q)/R_{0}^{2}>0$
have been introduced. Neglecting the running of the NC, we set~$q_{2}=0$
and therefore we have to solve~$q_{3}x^{-2}=1+q_{1}\ln x$ for~$x$.
The results are given by\begin{equation}
x=\frac{R_{\textrm{e}}}{R_{0}}=\sqrt{\frac{2q_{3}}{q_{1}\cdot W_{u}(\frac{2q_{3}}{q_{1}}e^{2/q_{1}})}},\label{eq:ReR0}\end{equation}
involving%
\begin{figure}[t]
\begin{center}\includegraphics[%
  clip,
  width=1.0\textwidth,
  keepaspectratio]{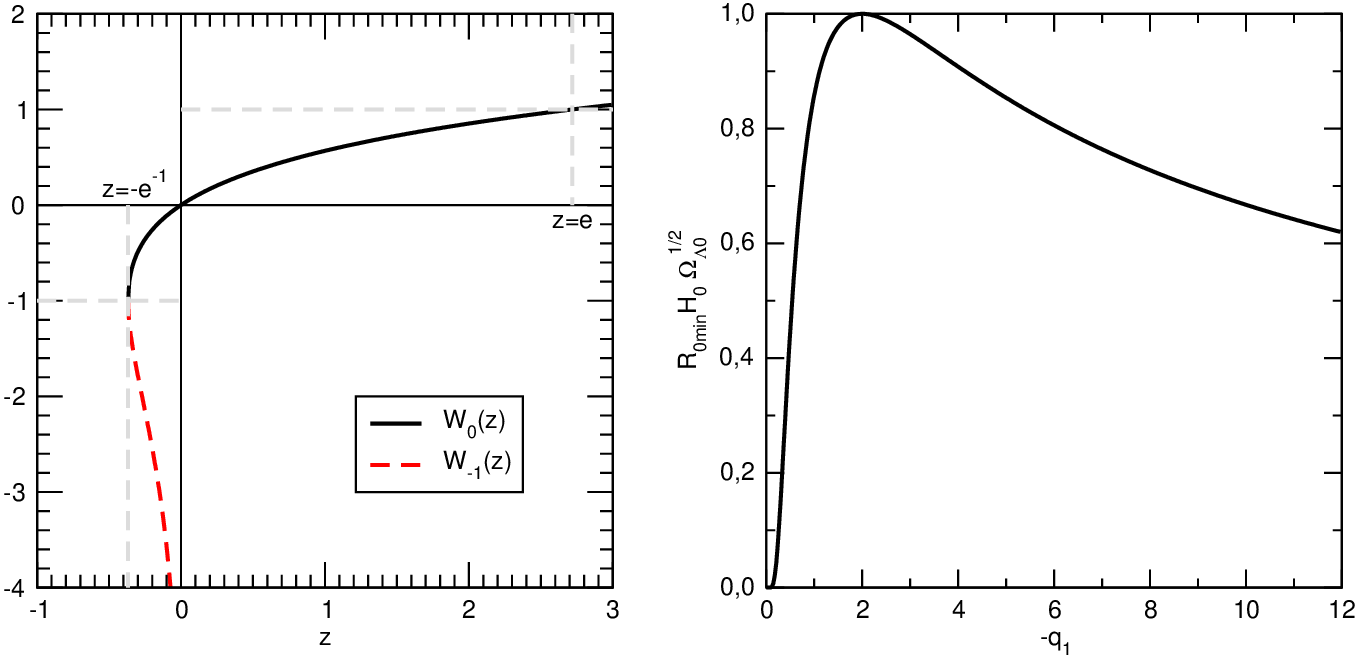}\end{center}

\caption{\label{cap:LambertW-R0min}Left: The real-valued branches of Lambert's
$W$-function~$W_{u}(z)$,~$u=0,-1$. Right: The lower bound~$R_{0\textrm{min}}$
of the initial event horizon radius~$R_{0}$ for negative~$q_{1}$.
For~$R_{0}<R_{0\textrm{min}}$ a final de~Sitter state does not
exist. }
\end{figure}
 Lambert's $W$-function~$W_{u}(z)$, which is the inverse function
of~$z=xe^{x}$. The index~$u$ denotes the different branches of
this function. Only for~$u=0,-1$ it takes on real values for real
arguments~$z>-e^{-1}$. Additionally,~$W_{-1}(z)$ is not real-valued
for~$z\ge0$, as can be seen in Fig.~\ref{cap:LambertW-R0min},
where both branches are plotted. From these properties of the $W$-function
we get for negative~$q_{1}$ the constraint~$q_{3}\le-\frac{q_{1}}{2}\exp(-\frac{2}{q_{1}}-1)$,
which implies an lower bound for the initial value~$R_{0}$ of the
horizon radius:\[
R_{0}\ge R_{0\textrm{min}}:=\sqrt{-\frac{2}{q_{1}\Omega_{\Lambda0}H_{0}^{2}}\exp\left(\frac{2}{q_{1}}+1\right)}.\]
If~$R_{0}$ is smaller than this minimal value, then Eq.~(\ref{eq:ReR0})
has no positive solutions and a final de~Sitter state does not exist.
For~$R_{0}=R_{0\textrm{min}}$ there is exactly one solution~$x=\exp(-\frac{1}{q_{1}}-\frac{1}{2})$,
for higher values~$R_{0}$ there are two solutions. %
\begin{figure}[t]
\begin{center}\includegraphics[%
  clip,
  width=0.70\textwidth,
  keepaspectratio]{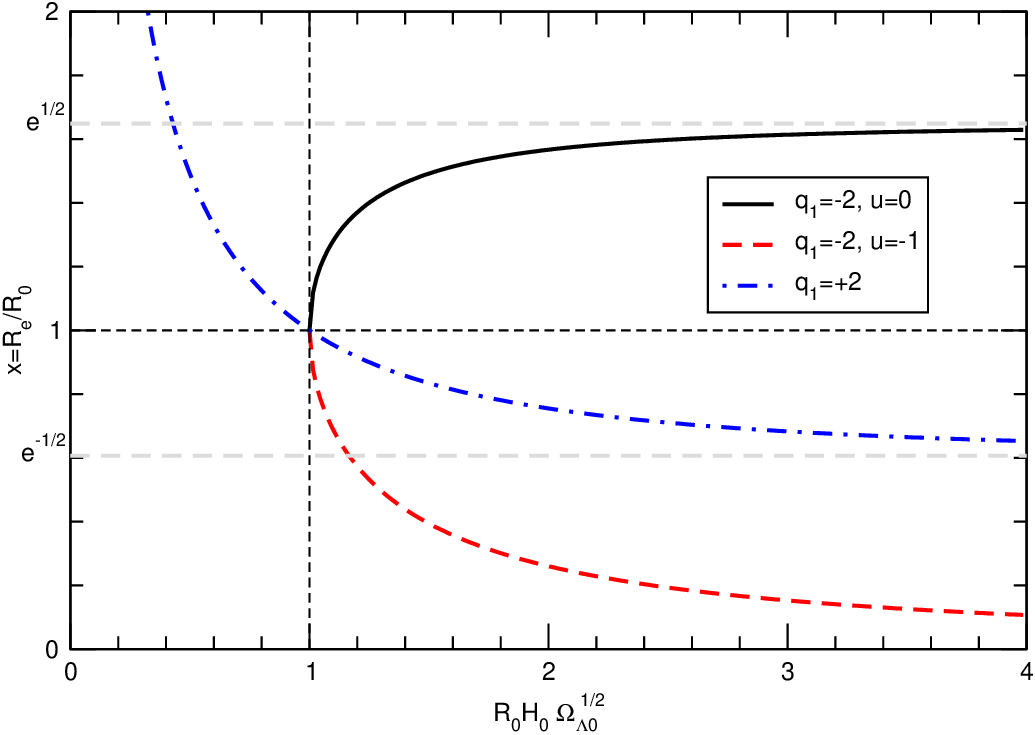}\end{center}

\caption{\label{cap:ReR0}The ratio~$x$ between the final event horizon
radius~$R_{\textrm{e}}$ and the initial radius~$R_{0}$ as a function
of~$R_{0}$, Eq.~(\ref{eq:ReR0}). Here~$q_{1}$ is set to~$\pm2$
and~$q_{2}=0$. All stable de~Sitter final states lie on the curve~$q_{2}=-2,\,\, u=0$.}
\end{figure}
In the case of a positive value of the parameter~$q_{1}$, the initial
value~$R_{0}$ must be smaller than~$1/\sqrt{H_{0}^{2}\Omega_{\Lambda0}}$.
Otherwise the final horizon radius~$R_{\textrm{e}}$ is smaller than
the initial one,~$x<1$. Both cases are plotted in Fig.~\ref{cap:ReR0}.

Since we have found several de~Sitter solutions, we have to study
the stability of these final states. Therefore, we write~$K_{0}\dot{F}$
as a function of~$K_{0}F=K_{0}(\dot{H}+(Q+1)H^{2})$,\[
K_{0}\dot{F}=q_{1}\left[H-\frac{1}{R_{0}}\exp\left(\frac{1-K_{0}F}{q_{1}}\right)\right],\]
where we used~$\dot{R}=RH-1$. In the final de~Sitter state we have~$K_{0}\dot{F}=0$
and~$R=R_{\textrm{e}}=1/H_{\textrm{e}}=\textrm{const}$. Near this
point we can neglect~$\dot{H}$ in the function~$F$ and replace~$H$
by $\sqrt{\frac{K_{0}F}{K_{0}(Q+1)}}$. For a stable solution it is
required that \[
\frac{\textrm{d}(K_{0}\dot{F})}{\textrm{d}(K_{0}F)}<0\]
in the final point, where~$K_{0}F=K_{0}(Q+1)H_{\textrm{e}}^{2}=q_{3}x^{-2}$.
With~$q_{3}$ and~$x$ from above, this yields the stability condition\[
\left[W_{u}(\kl{\frac{2q_{3}}{q_{1}}}e^{2/q_{1}})\right]^{-1}<-1,\]
implying that there are no stable de~Sitter solutions for positive
values of~$q_{1}$, because the $W$-function is positive. For negative~$q_{1}$
we get the condition\[
W_{u}(\kl{\frac{2q_{3}}{q_{1}}}e^{2/q_{1}})>-1,\]
which means that only the solution with $u=0$ is stable. This renders
the final event horizon radius~$R_{\textrm{e}}=R_{0}x$ unique.

\begin{figure}
\begin{center}\includegraphics[%
  clip,
  width=0.80\textwidth,
  keepaspectratio]{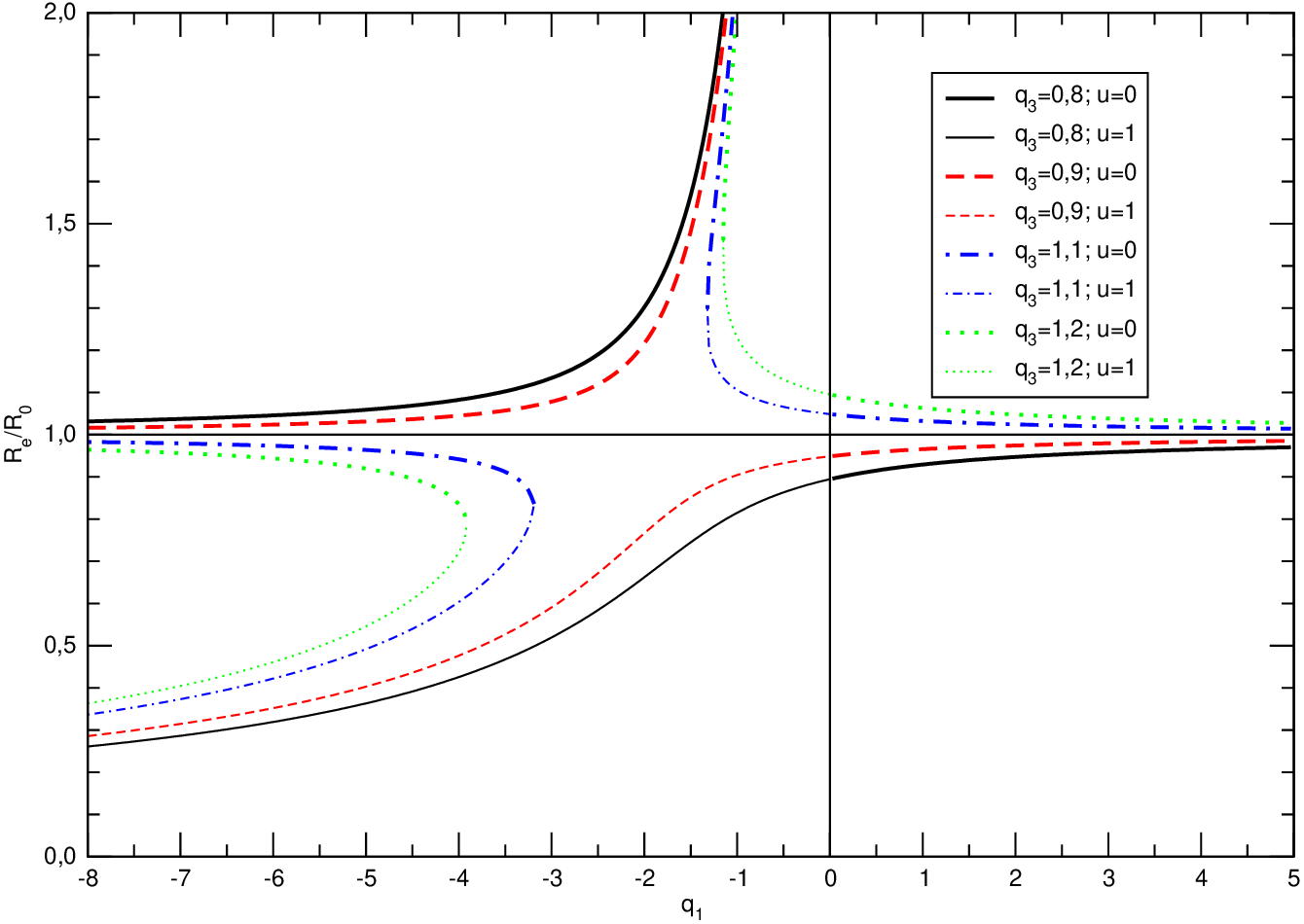}\end{center}

\caption{\label{cap:ReR0vonq1}This figure shows the ratio~$R_{\textrm{e}}/R_{0}$
between the final and the initial event horizon radius as a function
of~$q_{1}$. We plotted four cases for different values of~$q_{3}=1/(H_{0}^{2}R_{0}^{2}\Omega_{\Lambda0})$,
where the choice~$q_{3}=0,8;\,0,9$ corresponds to an initial radius~$R_{0}>1/(H_{0}\sqrt{\Omega_{\Lambda0}})$,
and~$q_{3}=1,1;\,1,2$ to~$R_{0}<1/(H_{0}\sqrt{\Omega_{\Lambda0}})$.
In the latter case there are no solutions for a certain range of values
of~$q_{1}$ as described by Eq.~(\ref{eq:q1Conditions}). The thick
lines show the ($u=0$)-branch of~$R_{\textrm{e}}/R_{0}$, and the
thin lines the ($u=-1$)-branch, respectively.}
\end{figure}

Finally, we take a closer look at the ratio~$R_{\textrm{e}}/R_{0}$
as a function of the mass parameter~$q_{1}$. For initial values~$R_{0}<1/(H_{0}\sqrt{\Omega_{\Lambda0}})$,
which means~$q_{3}>1$, there is a certain range of values of~$q_{1}$
where no solutions for~$R_{\textrm{e}}$ exist. This range is again
given by the requirement that the argument of the~$W$-function must
be greater than or equal to~$-e^{-1}$, leading to the conditions\begin{equation}
q_{1}\leq\frac{2}{W_{0}(-\frac{1}{eq_{3}})}\,\,\,\,\textrm{or}\,\,\,\, q_{1}\geq\frac{2}{W_{-1}(-\frac{1}{eq_{3}})}.\label{eq:q1Conditions}\end{equation}
In Fig.~\ref{cap:ReR0vonq1} the exclusion range for~$q_{1}$ is
obvious for~$q_{3}>1$. In the case that~$q_{1}$ lies above this
range, the unstable solution for~$R_{\textrm{e}}$ is reached first
during the future cosmic evolution. For~$q_{1}$ below this range,
the stable solution is nearer to the initial value~$R_{0}$ than
the unstable one, however, both solutions for~$R_{\textrm{e}}$ lie
below~$R_{0}$. Initial values~$R_{0}>1/(H_{0}\sqrt{\Omega_{\Lambda0}})$
(i.e.~$q_{3}<1$) lead to stable final states with~$R_{\textrm{e}}>R_{0}$
for all negative values of~$q_{1}$.

For initial values~$R_{0}$ and mass parameters~$q_{1}$, that do
not satisfy the above existence and stability conditions, the fate
of the universe will be {}``big crunch''-like or {}``big rip''-like%
\footnote{Recent analyses of future singularities and their properties can be
found in Refs.~\cite{BigRip-Analysis}.%
}, respectively. With these notations we mean that the scale factor~$a(t)$
or one of its derivatives~$H,q$ becomes singular in a finite time
in the future. There is one exception, that may occur when~$q_{1}$
and~$q_{2}$ are both positive, where the function~$F$ and thus
the Hubble scale~$H$ approach constant values, while the horizon
radius~$R$ goes to infinity, see Fig.~\ref{cap:q1plus2-q2plus01}.
Another property of the cosmic evolution for negative~$q_{1}$ is
that no big crunch may occur. This can be seen from the time-derivative
of the function~$K_{0}F$,\[
K_{0}\dot{F}=q_{1}\frac{\dot{R}}{R}=q_{1}\left(H-\frac{1}{R}\right),\,\,\, q_{1}<0,\]
which gets positive for~$H<0$ (big crunch), thus preventing a further
decrease of~$K_{0}F$ and the final collapse of the scale factor.

\section{\label{sec:DiscussNumeric}Numerical solutions}

Up to this section, we analysed the evolution equation~(\ref{eq:Hptglg})
for the scale factor analytically. Unfortunately, finding explicit
solutions seems to be rather difficult because of the strongly non-linear
form of the equation. Therefore, we solve it numerically, thereby
realising that the form given by Eq.~(\ref{eq:Hptglg}) is not directly
suitable for a numerical study because of the integral over the time~$t$
in the radius function~$R(t)$. This integral can be removed by differentiating
with respect to~$t$, leading to an ordinary differential equation
of the order three,\[
\left[\frac{\dot{a}}{a}+\frac{(q_{2}-q_{1})K_{0}\dot{F}}{(q_{1}-q_{2}K_{0}F)^{2}}\right]\cdot\exp\left[\frac{K_{0}F-1}{q_{1}-q_{2}K_{0}F}\right]-\frac{1}{R_{0}}=0,\]
which can be solved for the scale factor~$a(t)$ numerically. Note
that one has to check whether the numerical solutions are also solutions
of the original equation. This is not guaranteed since differentiating
Eq.~(\ref{eq:Hptglg}) possibly changes its set of solutions. Indeed,
we encounter such {}``false'' solutions in some cases.

Regarding recent observations, we get acceptable solutions only for~$\left|q_{1}\right|$
to be of the order~$1$, which means that the relevant mass scale~$m$
should be near~$\Lambda_{0}^{1/4}\sim10^{-3}\,\textrm{eV}$. Actually,
the only known particles with such a low mass are neutrinos. This
indicates that the influence of higher mass fields is suppressed,
or these fields have decoupled, respectively. Unfortunately, the simple
form of the RGEs~(\ref{eq:RGE-Lambda}) and~(\ref{eq:RGE-G}) cannot
account for a decoupling mechanism. Therefore, we assume in this work
that the mass scale~$m$ is low enough today, so that the solutions
are compatible with observations. However, note that at earlier cosmic
times high-mass fields~$m$ should be relevant.

Concerning the differential equation, we can fix several initial conditions
by using observational results. These are today's value of the Hubble
scale~$H_{0}=\frac{\dot{a}}{a}(t_{0})$ and the relative vacuum energy
density~$\Omega_{\Lambda0}$. Neglecting the spatial curvature~($\Omega_{k0}=-k/\dot{a}^{2}=0$)
and considering only dust (with an equation of state parameter~$Q=0,5$)
and the CC as relevant energy forms in the present-day universe, the
acceleration parameter~$q_{0}=\frac{\ddot{a}a}{\dot{a}^{2}}(t_{0})$
is determined by Eq.~(\ref{eq:Friedmann-3}): $q=\Omega_{\Lambda}(1+Q)+Q(\Omega_{k}-1)$.
Today's value of the horizon radius~$R_{0}$ is unknown, so we have
to estimate it. Since it should be the largest physical length scale
and the universe seems to be almost de~Sitter-like, we assume the
horizon radius to be around~$R_{0}\approx1,2\cdot H_{0}^{-1}$.

For the numerical treatment every dimensionful quantity is expressed
in terms of today's Hubble scale~$H_{0}$ (Hubble units). In a $\Lambda$CDM
universe with constant~$\Lambda$ and~$G$, the cosmic age is denoted
by~$t_{0}$. In our calculations we used the following numbers from
Ref.~\cite{PDG}:\[
H_{0}=1,5\cdot10^{-42}\,\textrm{GeV},\,\,\,\,\Omega_{\Lambda0}=0,73,\,\,\,\, t_{0}=13,7\,\textrm{Gyr}=0,99\cdot H_{0}^{-1},\]
\[
\Lambda_{0}=2,98\cdot10^{-47}\,\textrm{GeV}^{4},\,\,\,\,\Lambda_{0}^{\frac{1}{4}}\approx2,34\cdot10^{-3}\,\textrm{eV},\,\,\,\, G_{0}=(1,22\cdot10^{19}\,\textrm{GeV})^{-2}.\]

The first observation from the numerical solutions is, that for a
positive value of~$q_{1}$ the cosmic age decreases with respect
to the age~$t_{0}$ of the standard $\Lambda$CDM universe, whereas
for a negative~$q_{1}$ the age increases. Furthermore, the usually
small value of~$q_{2}$ leads to a negligible time variation of Newton's
constant~$G(t)$.

\begin{figure}
\begin{center}\includegraphics[%
  clip,
  width=0.85\textwidth,
  keepaspectratio]{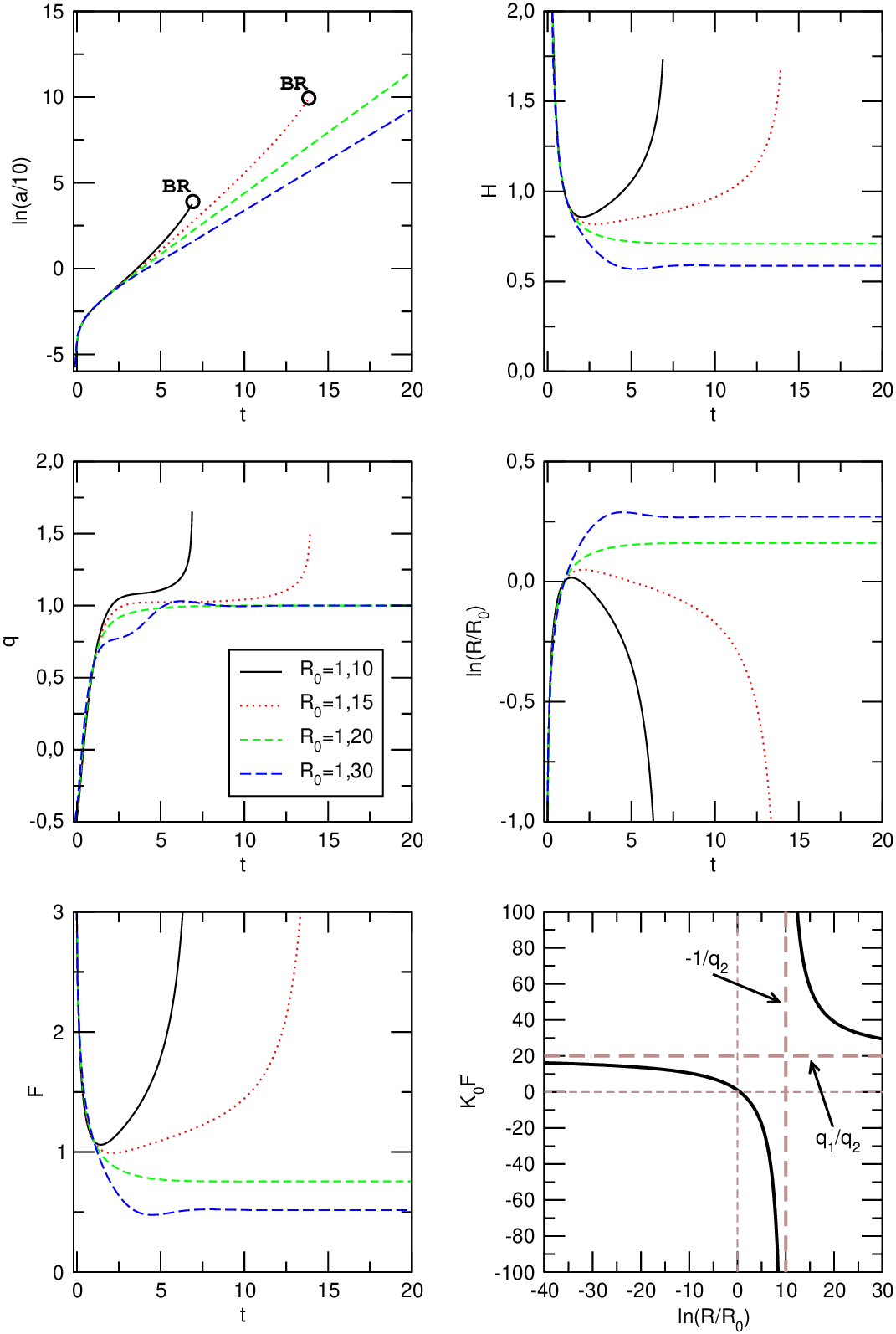}\end{center}

\caption{\label{cap:q1minus2-q2minus01}The cosmological evolution for the
parameter choice~$q_{1}=-2$ and~$q_{2}=-0,1$ and different values
of the initial event horizon radius~$R_{0}$. The fate of this universe
is either a stable de~Sitter state when choosing~$R_{0}=1,20;\,1,30$,
or a big rip~(\texttt{BR}) in the case~$R_{0}=1,10;\,1,15$. $K_{0}F$
is bounded from above. Nomenclature: Scale factor~$a$, Hubble scale~$H=\frac{\dot{a}}{a}$,
acceleration~$q=\frac{\ddot{a}a}{\dot{a}^{2}}$, event horizon radius~$R$
and its initial value~$R_{0}$. For the function~$K_{0}F$ see Eqs.~(\ref{eq:K0FofR}),~(\ref{eq:Friedmann-3}),
and~(\ref{eq:K0Def}), for the mass parameters~$q_{1},q_{2}$ see
Eqs.~(\ref{eq:q1Def}) and~(\ref{eq:q2Def}).}
\end{figure}

\begin{figure}
\begin{center}\includegraphics[%
  clip,
  width=0.85\textwidth,
  keepaspectratio]{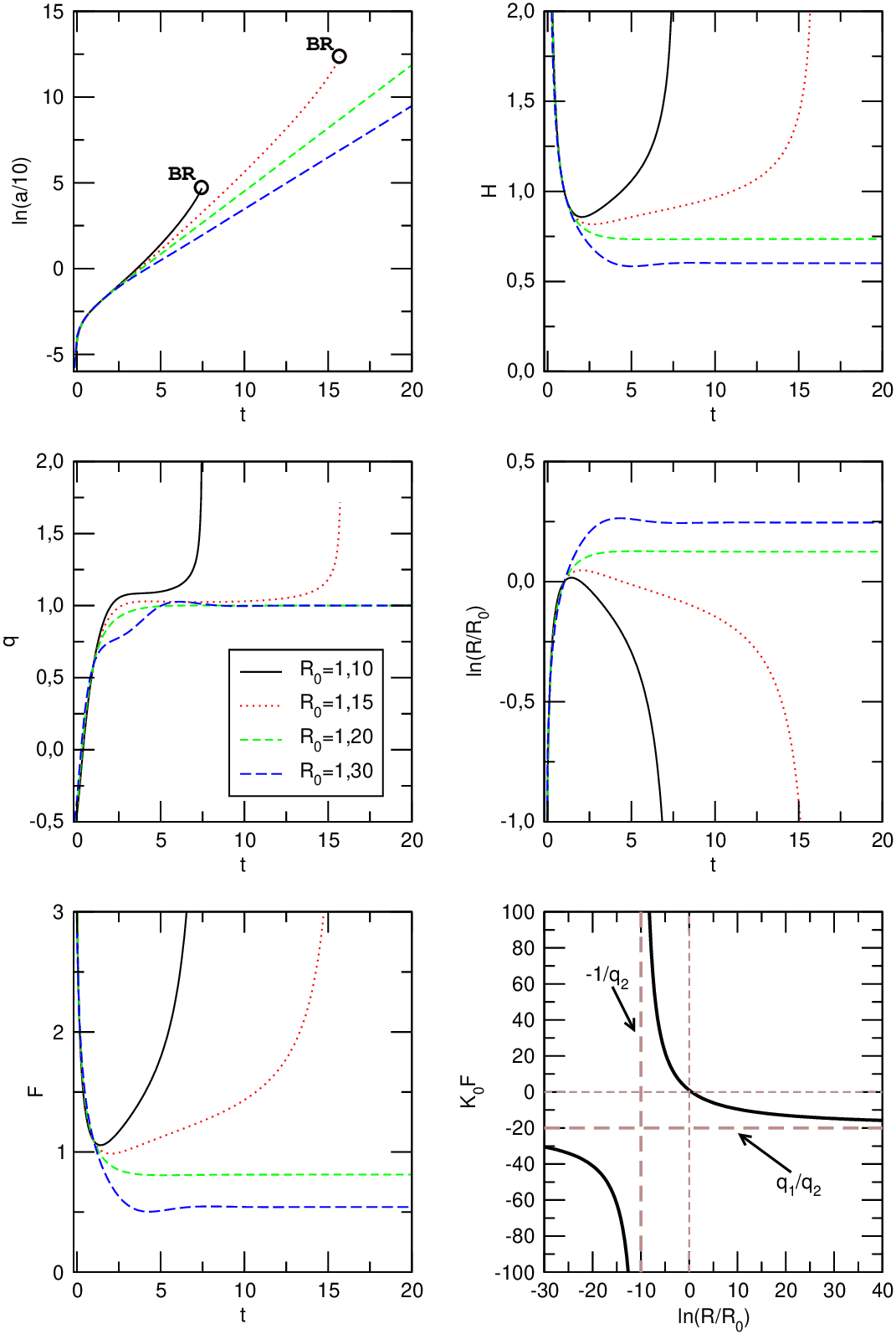}\end{center}

\caption{\label{cap:q1minus2-q2plus01}For the parameter choice~$q_{1}=-2$
and~$q_{2}=+0,1$ the cosmic evolution is not very different from
the case~$q_{2}=-0,1$ (Fig.~\ref{cap:q1minus2-q2minus01}). In
the future, there is either a stable de~Sitter state for~$R_{0}=1,20;\,1,30$,
or a big rip~(\texttt{BR}) when~$R_{0}=1,10;\,1,15$. $K_{0}F$
is bounded from below. Nomenclature: Scale factor~$a$, Hubble scale~$H=\frac{\dot{a}}{a}$,
acceleration~$q=\frac{\ddot{a}a}{\dot{a}^{2}}$, event horizon radius~$R$
and its initial value~$R_{0}$. For the function~$K_{0}F$ see Eqs.~(\ref{eq:K0FofR}),~(\ref{eq:Friedmann-3}),
and~(\ref{eq:K0Def}), for the mass parameters~$q_{1},q_{2}$ see
Eqs.~(\ref{eq:q1Def}) and~(\ref{eq:q2Def}).}
\end{figure}

\begin{figure}
\begin{center}\includegraphics[%
  clip,
  width=0.85\textwidth,
  keepaspectratio]{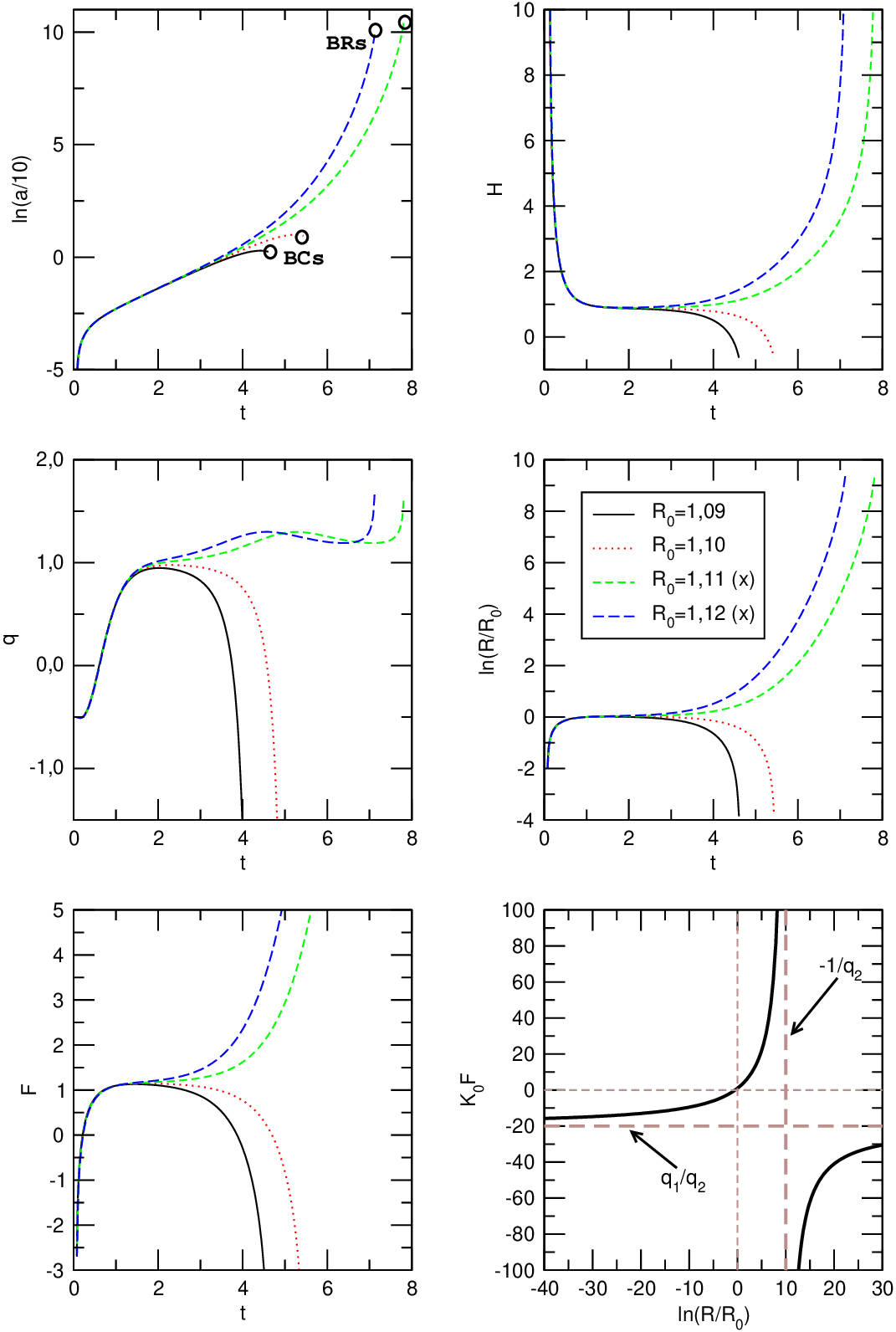}\end{center}

\caption{\label{cap:q1plus2-q2minus01}The cosmological evolution for different
values of today's horizon radius~$R_{0}$ for the case~$q_{1}=+2$
and~$q_{2}=-0,1$. The solutions for~$R_{0}=1,09;\,1,10$ exhibit
a big crunch~(\texttt{BC}), whereas the initial conditions~$R_{0}=1,11;\,1,10$
lead to a big rip~(\texttt{BR}). $K_{0}F$ is bounded from below.
The numerical solutions marked by~(\texttt{x}) are not compatible
with the main equation~(\ref{eq:Hptglg}), see Sec.~\ref{sec:DiscussNumeric}
for further details. Nomenclature: Scale factor~$a$, Hubble scale~$H=\frac{\dot{a}}{a}$,
acceleration~$q=\frac{\ddot{a}a}{\dot{a}^{2}}$, event horizon radius~$R$
and its initial value~$R_{0}$. For the function~$K_{0}F$ see Eqs.~(\ref{eq:K0FofR}),~(\ref{eq:Friedmann-3}),
and~(\ref{eq:K0Def}), for the mass parameters~$q_{1},q_{2}$ see
Eqs.~(\ref{eq:q1Def}) and~(\ref{eq:q2Def}).}
\end{figure}

\begin{figure}
\begin{center}\includegraphics[%
  clip,
  width=0.85\textwidth,
  keepaspectratio]{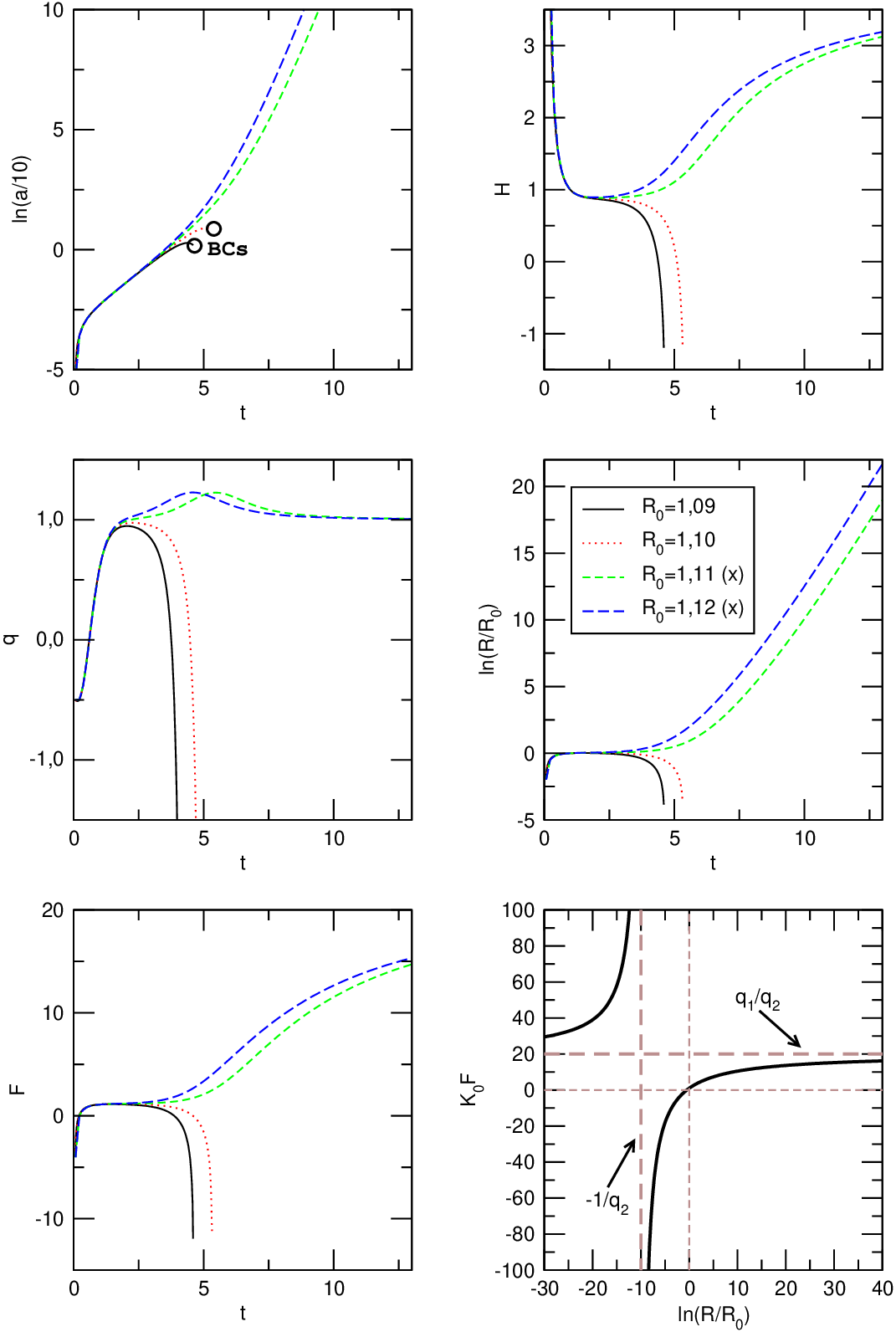}\end{center}

\caption{\label{cap:q1plus2-q2plus01}The cosmological evolution for different
values of today's horizon radius~$R_{0}$. Here, we choose~$q_{1}=+2$
and~$q_{2}=+0,1$. The solutions for~$R_{0}=1,09;\,1,10$ exhibit
a big crunch~(\texttt{BC}), where~$K_{0}F$ is unbounded from below.
For the initial conditions~$R_{0}=1,11;\,1,10$ the function~$F$
and the Hubble scale~$H$ approach a finite value, where the horizon
radius~$R$ diverges. The numerical solutions marked by~(\texttt{x})
are not compatible with the main equation~(\ref{eq:Hptglg}), see
Sec.~\ref{sec:DiscussNumeric} for further details. Nomenclature:
Scale factor~$a$, Hubble scale~$H=\frac{\dot{a}}{a}$, acceleration~$q=\frac{\ddot{a}a}{\dot{a}^{2}}$,
event horizon radius~$R$ and its initial value~$R_{0}$. For the
function~$K_{0}F$ see Eqs.~(\ref{eq:K0FofR}),~(\ref{eq:Friedmann-3}),
and~(\ref{eq:K0Def}), for the mass parameters~$q_{1},q_{2}$ see
Eqs.~(\ref{eq:q1Def}) and~(\ref{eq:q2Def}).}
\end{figure}

To show the characteristic future cosmic evolution, we investigate
four cases in more detail, which result from the parameter choices~$q_{1}=\pm2$
and~$q_{2}=\pm0,1$. Note that due to the suppression by the Planck
scale, the realistic value of~$q_{2}$ should be much lower than~$\pm0,1$.
Here, we used a large value for~$q_{2}$ to show the differences
due to the sign of~$q_{2}$. Figures~\ref{cap:q1minus2-q2minus01}--\ref{cap:q1plus2-q2plus01}
show the numerical results for different values of the initial radius~$R_{0}$
of the event horizon. The graphs in each of the four figures illustrate
the scale factor~$a(t)$, the Hubble scale~$H(t)$, the acceleration~$q(t)$,
the event horizon radius~$R(t)$, and~$F(t)$ as functions of the
cosmic time~$t$, respectively. The last graph displays~$K_{0}F$
as a function of the radius~$R/R_{0}$.

In section~\ref{sec:DiscussAnalyt} we discussed the evolution equation
analytically, and we found several conditions for the existence of
stable de~Sitter final states. These properties are also shown by
the the numerical results. For negative values of~$q_{1}$, we observe
only {}``big rip''-like solutions and de~Sitter final states, whereas
for positive~$q_{1}$, a big crunch may also occur, and all de~Sitter
states are unstable. Note that the catastrophic events, the {}``big
rip'' and the {}``big crunch'', usually involve a high gravitational
strength, implying that our calculations may not be reliable near
these singular points. For positive values of~$q_{1}$ and~$q_{2}$
(see Fig.~\ref{cap:q1plus2-q2plus01}), we have not observed any
{}``big rip''-like solutions. Then the final state may be either
a {}``big crunch'' or a forever expanding universe, where the Hubble
scale approaches a finite positive value, but the event horizon radius~$R$
goes to infinity. This is a contradiction, because an asymptotically
constant Hubble scale~$H>0$ implies a finite event horizon radius~$R\approx H^{-1}$
in the far future, which is not the case here. Obviously, this numerical
solution is not a solution of the original equation~(\ref{eq:Hptglg}).
For~$q_{1}>0$ and~$q_{2}<0$ (see Fig.~\ref{cap:q1plus2-q2minus01})
the {}``big rip''-like events in the numerical solutions occur at
a finite and large value of the horizon radius~$R$. Again, this
behaviour is not compatible with the vanishing of the horizon radius
at such an event. Therefore, we can reject these numerical solutions,
too.

\section{\label{sec:Conclusions}Conclusions}

We investigated a cosmological model, where the CC and the NC are
determined by renormalisation group equations, which emerge from QFT
on curved space-time. By choosing the inverse radius of the cosmological
event horizon as the renormalisation scale, we get time-dependent
constants. Because of this, the evolution equation for the cosmic
scale factor becomes more complicated than in standard $\Lambda$CDM-cosmology,
leading to cosmological solutions with several very different future
final states of the universe. We found {}``big crunch''-like and
{}``big rip''-like solutions, but also stable de~Sitter final states.
Which of these states will be realised depends crucially on the field
masses in the renormalisation group equations, and on the initial
value of today's radius of the event horizon. In this context, we
derived some conditions on the existence of stable de~Sitter states.
Furthermore, the cosmic evolution was analysed numerically for different
values of the field masses. For a realistic cosmic behaviour, we have
to require that the masses in the RGEs should be quite low, implying
the need for some suppression or decoupling mechanism for high-mass
fields. However, such a mechanism for the CC and the NC has not been
found yet~\cite{DecLamG}. This also restricts the main focus of
this work to the future behaviour of the universe, because for the
cosmic evolution at early times quantum fields with high masses should
be taken into account. Moreover, the regimes with a high gravitational
strength need a deeper investigation, because such conditions are
given not only at early times, but also at the singularities in the
future. Finally, we conclude, that the specific choice of the renormalisation
scale in this work leads to cosmological solutions, that may become
singular in a finite time without introducing exotic forms of matter.

\section*{Acknowledgements}

I would like to thank M. Lindner for useful comments and discussions.
This work was supported by the {}``Sonderforschungsbereich 375 für
Astroteilchenphysik der Deutschen Forschungsgemeinschaft'', and I
wish to thank the Freistaat Bayern for a {}``Promotionsstipendium''.

\newpage

\end{document}